%% file: bare_jrnl.tex
\documentclass[conference,10pt]{IEEEtran}
\IEEEoverridecommandlockouts

\input{macros}
\renewcommand{\baselinestretch}{0.99} 
\begin{document}

\input{acronyms}

\title{Performance Analysis of Cell-Free Massive MIMO under Imperfect LoS Phase Tracking}

% \author{\IEEEauthorblockN{Noor~Ul~Ain,~Lorenzo~Miretti,~Renato~L.~G.~Cavalcante~and~S{\l}awomir~Sta{\'n}czak}}
\author{\IEEEauthorblockN{Noor~Ul~Ain\IEEEauthorrefmark{1},
Lorenzo~Miretti\IEEEauthorrefmark{2},
Renato~L.~G.~Cavalcante\IEEEauthorrefmark{3},
and
S{\l}awomir~Sta{\'n}czak\IEEEauthorrefmark{1}\IEEEauthorrefmark{3}\thanks{The authors acknowledge funding by the Bundesministerium für Forschung, Technologie und Raumfahrt (BMFTR, German Federal Ministry of Research, Technology and Space) project xG-RIC under Grants 16KIS2429K and 16KIS2434, and by the 6G-MIRAI project, which has received funding from the Smart Networks and Services Joint Undertaking (SNS JU) under the European Union’s Horizon Europe research and innovation programme, Grant Agreement No. 10119236. Additionally, the authors acknowledge financial support from the Federal Ministry for Digital and Transport of Germany under Grant 19OI23001A. Views and opinions expressed are, however, those of the author(s) only and do not necessarily reflect those of the European Union or the SNS JU (granting authority). Neither the European Union nor the granting authority can be held responsible for them. Most contributions of L. Miretti were made while he was with the Fraunhofer Heinrich Hertz Institute (HHI), Berlin, Germany.}}
\IEEEauthorblockA{\IEEEauthorrefmark{1}Technical University of Berlin, Berlin, Germany, noor.ul.ain@tu-berlin.de}
\IEEEauthorblockA{\IEEEauthorrefmark{2}Ericsson Research, Herzogenrath, Germany,
lorenzo.miretti@ericsson.com}
\IEEEauthorblockA{\IEEEauthorrefmark{3}Fraunhofer Heinrich Hertz Institute, Berlin, Germany {\{renato.cavalcante,slawomir.stanczak\}@hhi.fraunhofer.de}}
}
\maketitle

\begin{abstract}
We study the impact of imperfect line-of-sight (LoS) phase tracking on the uplink performance of cell-free massive MIMO networks. Unlike prior works that assume perfectly known or completely unknown phases, we consider a realistic regime where LoS phases are estimated with residual uncertainty due to hardware impairments, mobility, and synchronization errors. To this end, we propose a Rician fading model where LoS components are rotated by imperfect phase estimates and attenuated by a deterministic \textit{phase-error penalty factor}.

We derive a linear MMSE channel estimator that accounts for statistical phase errors and unifies prior results, reducing to the Bayesian MMSE estimator when phase is perfectly known and to a zero-mean model when no phase information is available. To address the non-Gaussian setting, we introduce a virtual uplink model that preserves second-order statistics of channel estimation, enabling the derivation of tractable virtual centralized and distributed MMSE beamformers. To ensure fair assessment of network performance, we apply these virtual beamformers to the operational uplink model that reflects the actual physical channel and compute the spectral efficiency bounds available in the literature.

Numerical results show that our framework bridges idealized assumptions and practical tracking limitations, providing rigorous performance benchmarks and design insights for 6G cell-free networks.
\end{abstract}
\begin{IEEEkeywords}
cell-free MIMO, line-of-sight, MMSE, channel estimation, phase noise, imperfect CSI
\end{IEEEkeywords}

\section{Introduction}\label{sec:intro}

Cell-free massive MIMO (mMIMO) has emerged as a promising wireless architecture to meet the demands of beyond-5G and 6G networks. By deploying a large number of distributed access points (APs) that coherently serve all users, cell-free systems eliminate inter-cell boundaries, improve uniformity of service, and provide high spectral and energy efficiency~\cite{Demir2021,ngo_2017,interdonato2019}. Considering the high likelihood of LoS propagation in cell-free networks due to dense deployment regimes, many recent works have studied the performance of such networks under LoS channel conditions~\cite{ain2025,polegre2020channel,Ozdogan2019,9740487,wang2021,wang2024,Ngo2018}. As shown in~\cite{ain2025}, LoS components play a significant role in the performance of distributed architectures when optimal beamforming schemes are employed. 

A key challenge, however, lies in the acquisition of reliable channel state information (CSI) when LoS phases fluctuate due to oscillator drift, user mobility, or synchronization errors. Such fluctuations lead to residual phase uncertainty, even after tracking, and can cause non-negligible performance degradation.

Existing studies typically analyze cell-free systems under two limiting cases of LoS phase knowledge. On one hand, studies such as~\cite{wang2024,ain2025} and phase-aware case in~\cite{Ozdogan2019,wang2021} assume perfect knowledge of the LoS phases, which enables tractable analysis but overlooks practical estimation errors. On the other hand, studies~\cite{wang2021,Ozdogan2019} also consider the case of completely unknown LoS phases, thereby discarding valuable deterministic information inherent in the slowly varying LoS components. However, as also pointed out in~\cite{wang2021,Ozdogan2019}, the practical system performance fall in between these two scenarios. Furthermore, the performance evaluation of LoS-aware cell-free systems has largely been restricted to maximum-ratio (MR) combining~\cite{wang2024,wang2021,Ozdogan2019}. While analytically convenient, MR combining does not fully exploit the beamforming potential of large-scale cell-free networks. Complementary studies~\cite{hardwareimpairements,phasenoise_frequencyselective,7491244} have investigated phase noise effects in Rayleigh fading channels, that is, without explicit modeling of LoS components. These simplifications limit our understanding of the impact of partial LoS phase knowledge on channel estimation and beamforming in cell-free systems.

To bridge this gap, we consider a realistic intermediate regime where LoS phases are imperfectly tracked. Unlike prior works that assume either perfectly known or completely unknown phases, we explicitly model partial phase knowledge by introducing a Rician fading model in which the deterministic LoS component is rotated by a noisy phase estimate and attenuated by a deterministic penalty factor that depends on the tracking accuracy. The residual phase error is modeled as a bounded random variable, reflecting the fact that hardware-induced oscillator drifts and synchronization offsets are typically limited in range. Although hardware-level analyses commonly assume Gaussian phase noise models, the uniform model provides an analytically tractable worst-case approximation that bounds the effective phase uncertainty.

Building on this model, we derive a linear minimum mean-squared error (linear MMSE) channel estimator that remains tractable in non-Gaussian settings and unifies prior results in the literature. The estimator coincides with the Bayesian MMSE estimator under perfect phase knowledge~\cite{ain2025, Ozdogan2019,wang2024} and with the zero-mean model when phases are entirely unknown~\cite{Ozdogan2019,wang2021}, thereby continuously interpolating between these two extremes. We further extend the framework to centralized and distributed MMSE beamforming by introducing a virtual uplink model that embeds estimation error covariance into independent additive noise, enabling pseudo closed-form solutions. Numerical evaluations show that even coarse phase tracking yields substantial performance gains compared to the no phase knowledge model. Overall, the proposed approach unifies phase-aware estimation and beamforming design under a common framework suitable for realistic 6G cell-free massive MIMO deployments.

\textit{Notation:} Lower and upper case bold letters are used for vectors and matrices, respectively, while calligraphic letters are used for sets. The transpose and Hermitian transpose is denoted as $(.)^{\mathsf{T}}$ and $(.)^{\mathsf{H}}$ respectively. A block-diagonal matrix with matrices $\bm{D}_1,\dots,\bm{D}_N$ on its diagonal is denoted as $\mathrm{diag}(\bm{D}_1,\dots,\bm{D}_N)$. The expectation and variance of a random quantity $A$ are denoted by $\mathbb{E}[A]$ and $\mathbb{V}[A]$ and their conditional version given a random quantity $B$ are denoted by $\mathbb{E}[A|B]$ and $\mathbb{V}[A|B]$ respectively. We denote by $\mathbb{R}_+$ the set of all positive real numbers i.e. $\mathbb{R}_+:=]0,\infty[$.

\section{System and Channel Model}\label{sec:channel}
We consider a user-centric cell-free \ac{mmimo} network with $L$ \acp{ap}, indexed by $\mathcal{L} := \{1, \dots, L\}$, jointly serving $K$ users indexed by $\mathcal{K}:= \{1, \dots, K\}$. The positions are fixed and arbitrarily distributed across the service area. All \acp{ap} are connected to a \ac{cpu} via an error-free fronthaul network~\cite{ngo_2017}. Each \ac{ap} is equipped with $N$ antennas, while each user has a single antenna. 

To capture LoS propagation, we adopt a spatially correlated Rician fading model that incorporates phase shifts. Under the standard block-fading assumption, the channel remains time-invariant and frequency-flat within each coherence block of $\tau_c$ symbols. Across coherence blocks, the channel evolves according to a stationary and ergodic random process. However, since our results only depend on its first-order distribution, we do not explicitly model higher-order statistics such as correlation across blocks.

\subsection{Rician Channel under Imperfect Phase Tracking}
In an arbitrary coherence block, let $\bm{h}_{k,l} \in \mathbb{C}^N$ denote a realization of the channel vector between the user $k \in \mathcal{K}$ and the $N$ antennas of \ac{ap} $l \in \mathcal{L}$. More specifically, the channel is modeled as $(\forall k\in\mathcal{K})(\forall l\in\mathcal{L})$
\begin{equation} \label{eq:channel}
    \bm{h}_{k,l} = \bm{\bar{h}}_{k,l} e^{j\theta_{k,l}} + \bm{\tilde{h}}_{k,l}, 
    \quad \bm{\tilde{h}}_{k,l} \sim \mathcal{CN}(\mathbf{0}, \mathbf{R}_{k,l}),
\end{equation}
where $\bm{\bar{h}}_{k,l} \in \mathbb{C}^N$ is the deterministic spatial signature of the LoS component, and $\mathbf{R}_{k,l} \in \mathbb{C}^{N \times N}$ is the spatial covariance matrix of the \ac{nlos} components $\bm{\tilde{h}}_{k,l}$, and $\theta_{k,l}\sim \mathcal{U}[-\pi,\pi]$ is the phase of the LoS component.  For each realization of $\theta_{k,l}$, an estimate \begin{equation} \label{eq:phase}
\hat{\theta}_{k,l}=\theta_{k,l}-\varepsilon_{k,l}\end{equation}
is observed, where $\varepsilon_{k,l}$ denotes the estimation error. 

Existing works such as~\cite{Ozdogan2019,wang2021} consider special cases of the above model under the two extremes of either perfectly known or completely unknown LoS phases. In the particular assumption of unknown phase, the underlying motivation is that the LoS phase $\theta_{k,l}$ often varies more rapidly than the large scale fading components $\bm{\bar{h}}_{k,l}$ and $\bm{R}_{k,l}$, which makes accurate estimation of $\theta_{k,l}$ challenging, particularly in high mobility scenarios or when the AP has a single antenna~\cite{Ozdogan2019}.

However, as noted in~\cite{Ozdogan2019,wang2021}, practical systems operate between these extremes. In fact, the LoS phase often varies more slowly than the \ac{nlos} components, whose dynamics are dominated by delay and Doppler spreads. For example, recent work~\cite{wang2024} claims that the LoS phase may remain approximately constant across multiple coherence blocks, particularly in the frequency domain. Therefore, with our model we focus on the intermediate case where the LoS phase $\theta_{k,l}$ varies more slowly than the NLoS components $\bm{\tilde{h}}_{k,l}$, but still fast enough such that it cannot be tracked with desired accuracy. We denote by $\bm{\hat{\theta}}_{l}=[\hat{\theta}_{1,l},\dots, \hat{\theta}_{K,l}]^\mathsf{T}\in\mathbb{C}^{ 1\times K}$ and $\bm{\hat{\theta}}=[\bm{\hat{\theta}}_1,\dots, \bm{\hat{\theta}}_L]\in\mathbb{C}^{ L\times K}$ the phase estimates for all channels of AP $l$ and the network respectively. 

The phase estimates $\bm{\hat{\theta}}$ are assumed to be available as common knowledge across the network. The phase estimation error $\varepsilon_{k,l}$ in \eqref{eq:phase} is modeled as uniformly distributed, $\varepsilon_{k,l}\sim\mathcal{U}[-\delta,\delta]$, with $0\leq\delta\leq\pi$, and it is assumed independent of the \ac{nlos} component.  Under these assumptions, the conditional distribution of the channel admits simple expressions for the first and second moments, summarized below (the proof is provided in the Appendix~\ref{app:lemma}).
% (the proof is omitted due to the space limitation and will be provided elsewhere).

\begin{lemma}\label{lem:moments}
Fix $\delta\in [0,\pi]$. Let $\varepsilon_{k,l}\sim\mathcal{U}[-\delta,\delta]$ and let $\bm{\tilde{h}}_{k,l}$ be independent of $\hat\theta_{k,l}$. Define
\begin{equation}
    \rho_{\mathrm{error}} := \mathbb{E}\!\left[e^{j\varepsilon_{k,l}}\right] =
    \begin{cases}
        1, & \delta=0, \\[4pt]
        \dfrac{\sin(\delta)}{\delta}, & 0<\delta\leq \pi.
    \end{cases}
\end{equation}
Then, we have
\[
    \mathbb{E}[\mathbf{h}_{k,l}\mid \hat{\theta}_{k,l}]= \bm{\bar{h}}_{k,l}\, e^{j\hat{\theta}_{k,l}}\, \rho_{\mathrm{error}}\quad\text{and},\]
   \[ \mathbb{V}[\mathbf{h}_{k,l}\mid \hat{\theta}_{k,l}]= \mathbf{R}_{k,l} + \bm{\bar{h}}_{k,l}\bm{\bar{h}}_{k,l}^{H}(1-\rho_{\mathrm{error}}^{2})=:\bm{\Sigma}_{k,l}.
\]
\end{lemma}

% Proof: see Appendix~\ref{app:moments}
\begin{remark}
When the phase error $\varepsilon_{k,l}\sim\mathcal{U}[-\delta,\delta]$ vanishes as $\delta \to 0$, corresponding to perfect phase tracking, the factor $\rho_{\mathrm{error}}$ approaches one. Conversely, when the phase error becomes maximally uncertain as $\delta \to \pi$, reflecting completely unknown phases, the factor $\rho_{\mathrm{error}}$ tends to zero. In general, imperfect tracking rotates the deterministic LoS component $\bm{\bar{h}}_{k,l}$ by $\hat{\theta}_{k,l}$, and it reduces its magnitude by the factor $\rho_{\mathrm{error}}$, which we refer to as the \emph{phase noise penalty}.
\end{remark}

\section{Channel Estimation} 

For pilot signaling, the network employs $\tau_p$ out of $\tau_c\geq\tau_p$ symbols in a coherence block to transmit $\tau_p$ mutually
orthogonal pilot sequences. We assume that the number $K$ of users is large such that $\tau_p \ll K$, thus some pilots must be shared by more than one user. We denote by \( \mathcal{P}_k \) the set of users sharing the same pilot sequence as user \( k \). The received signal at \ac{ap} $l$, after decorrelating with respect to the pilot sequence indexed as $t_k\in \{1,\ldots,\tau_p\}$ assigned to user $k\in \mathcal{K}$ , is given by~\cite{Demir2021}

\begin{equation}\label{eq:ul_pilot_signal}
\bm{y}_{t_k,l}^{\mathrm{pilot}}=\sum_{i\in\mathcal{P}_k}\sqrt{\eta_i}\tau_p\bm{h}_{i,l}+\bm{n}_{t_k,l},
\end{equation}
where $\eta_i\in\mathbb{R}_+ $ is the pilot transmit power of user~$i$, \( \bm{h}_{i,l} \) is the channel from user~$i$ to \ac{ap}~$l$ and \( \bm{n}_{t_k,l} \sim \mathcal{CN}(0, \sigma^2 \tau_p \bm{I}_N) \) is the additive Gaussian noise at the receiver side.

Under the channel model in \eqref{eq:channel}, the channel $\bm{h}_{k,l}$ and the measured pilot signal $\bm{y}_{t_k,l}^{\mathrm{pilot}}$ (together with the phase estimates $\bm{\hat{\theta}}_{l}$) are not jointly Gaussian. Hence, the Bayesian MMSE estimator $\mathbb{E}[\bm{h}_{k,l}\!\mid\!\bm{y}_{t_k,l},\bm{\hat{\theta}}_{l}]$ does not necessarily admit a simple closed-form expression and requires integration over the phase distribution. To obtain a tractable channel estimation solution in this non-Gaussian setting, we derive a conditionally \emph{linear} MMSE estimator in $\bm{y}_{t_k,l}^{\mathrm{pilot}}$ given $\bm{\hat{\theta}}_{l}$, which minimizes the MSE among all linear functions of $\bm{y}_{t_k,l}^{\mathrm{pilot}}$ for a given phase estimate~\cite[Sec.~12.3]{kay_1993}. This choice is crucial as it retains analytical tractability in non-Gaussian settings and, as shown in Corollaries~\ref{cor:perfect} and~\ref{cor:unknown}, it yields the classical MMSE estimators as limiting cases.

\begin{definition}[Conditionally Linear MMSE Estimator~\cite{kay_1993}]\label{def:lmmse}
For random vectors $\bm{x}$ and $\bm{y}$, and a conditioning random quantity $\bm{z}$ (e.g., phase estimates $\hat{\boldsymbol{\theta}}$), the \emph{conditionally linear MMSE estimator} of $\bm{x}$, given $\bm{y}$ and $\bm{z}$, is the affine function of $\bm{y}$ that minimizes the conditional mean-squared error $\mathbb{E}\big[\lVert \bm{x}-\bm{\hat{x}}\rVert^2\!\mid\bm{z}\big]$. It is given by
\[
\hat{\bm{x}}(\bm{y}\mid\bm{z})
= \mathbb{E}[\bm{x}\mid\bm{z}]
+ \bm{R}_{xy|\bm{z}}\bm{R}_{yy|\bm{z}}^{-1}
\big(\bm{y}-\mathbb{E}[\bm{y}\mid\bm{z}]\big),
\]
where $\bm{R}_{xy|\bm{z}}$ and $\bm{R}_{yy|\bm{z}}$ are the conditional covariance matrices, and $\bm{R}_{yy|\bm{z}}$ is assumed to be full rank.
\end{definition}

\begin{proposition}[Conditionally Linear MMSE channel estimator under imperfect phase tracking]\label{prop:LMMSE}
Using the measured signal $\bm{y}_{t_k,l}^{\mathrm{pilot}}$ and conditioning on the phase estimates $\bm{\hat{\theta}}_{l}$, the conditionally linear MMSE estimator of $\bm{h}_{k,l}$ is given by (the proof is provided in the Appendix~\ref{app:prop})
% (the proof is omitted due to space limitations and will be provided elsewhere) 
$(\forall k\in\mathcal{K})(\forall l\in\mathcal{L})$
\begin{equation}\label{eq:estimator}
\bm{\hat{h}}_{k,l}=\bm{\bar{h}}_{k,l}e^{j\hat{\theta}_{k,l}}\rho_{\mathrm{error}} 
+ \sqrt{\eta_k}\bm{\Sigma}_{k,l} \bm{\Psi}_{t_k,l} \Big( \bm{y}_{t_k,l}^{\mathrm{pilot}} - \bm{\bar{y}}_{t_k,l}\Big),
\end{equation}
\begin{align*}
\text{where}\quad\bm{\Psi}_{t_k,l}&=\left(\sum_{i\in\mathcal{P}_k}\eta_i\tau_p \bm{\Sigma}_{i,l} +\sigma^2\bm{I}_N\right)^{-1}, \\
\bm{\bar{y}}_{t_k,l}&=\sum_{i\in\mathcal{P}_k}{\sqrt{\eta_i}\tau_p}\bm{\bar{h}}_{i,l}  e^{j\hat{\theta}_{i,l}}\rho_{\mathrm{error}},
\end{align*}
and $\bm{\Sigma}_{k,l}$ is the conditional covariance matrix of channel $\bm{h}_{k,l}$, given in Lemma~\ref{lem:moments}. Furthermore, the estimation error $\bm{\xi}_{k,l}=\bm{{h}}_{k,l}-\bm{\hat{h}}_{k,l}$ has zero mean and the covariance matrix given by
\begin{equation}\label{eq:errorcovariance}
\bm{C}_{k,l}=\bm{\Sigma}_{k,l}-\eta_k\tau_p\bm{\Sigma}_{k,l}\bm{\Psi}_{t_k,l}\bm{\Sigma}_{k,l}.
\end{equation}
The local channel estimate $\bm{\hat{h}}_{k,l}$ and the estimation error $\bm{\xi}_{k,l}$ are uncorrelated random vectors satisfying $(\forall k\in \mathcal{K})(\forall l\in \mathcal{L})$
$ \mathbb{E}[\bm{\hat{h}}_{k,l}\mid{\hat{\theta}}_{l}]=\bm{\bar{h}}_{k,l} e^{j\hat{\theta}_{k,l}}\rho_{\mathrm{error}},$
$ \mathbb{V}[\bm{\hat{h}}_{k,l}\mid{\hat{\theta}}_{l}]=\bm{\Sigma}_{k,l}-\bm{C}_{k,l},$
$ \mathbb{E}[\bm{\xi}_{k,l}\mid{\hat{\theta}}_{l}]=\bm{0}$ and $ \mathbb{V}[\bm{\xi}_{k,l}\mid{\hat{\theta}}_{l}]=\bm{C}_{k,l}$.
%Proof. See Appendix \ref{app:lmmse}. 
\end{proposition}
We assume that all statistical parameters such as $\bm{\bar{h}}_{k,l}$, $\bm{R}_{k,l}$, $\sigma^2$ and $\rho_{\mathrm{error}}$ are known across the network. The resulting estimator $\bm{\hat{h}}_{k,l}$ is affine in the received signal $\bm{y}_{t_k,l}$, but nonlinear in the phase estimate $\bm{\hat{\theta}}_{l}$.

For convenience, we denote by 
$\bm{\hat{H}}_{l}=[\bm{\hat{h}}_{1,l},\dots,\bm{\hat{h}}_{K,l}]\in\mathbb{C}^{N\times K}$ the local estimate of the local channel $\bm{{H}}_{l}=[\bm{{h}}_{1,l},\dots, \bm{{h}}_{K,l}]\in\mathbb{C}^{N\times K}$ of \ac{ap}~$l$ and by $\bm{\hat{H}}=[\bm{\hat{H}}_{1}^{\mathsf{T}},\dots,\bm{\hat{H}}_{L}^{\mathsf{T}}]^{\mathsf{T}}\in\mathbb{C}^{LN\times K}$ the global estimate of the global channel $\bm{{H}}=[\bm{{H}}_{1}^{\mathsf{T}},\dots,\bm{{H}}_{L}^{\mathsf{T}}]^{\mathsf{T}}\in\mathbb{C}^{LN\times K}$.

The next two corollaries present the limiting behaviour of the linear MMSE estimator in Proposition~\ref{prop:LMMSE}, as the phase tracking error vanishes or becomes maximal.
\begin{corollary}{(Perfect phase knowledge)}\label{cor:perfect}
For $\varepsilon_{k,l}\sim\mathcal{U}[-\delta,\delta]$, as $\delta\to 0$ (i.e., $\rho_{\mathrm{error}}\to 1$), the joint distribution of $\bm{y}_{t_k,l},$ and $\bm{\bar{h}}_{k,l}$ becomes Gaussian, and the linear MMSE estimator in Proposition~\ref{prop:LMMSE} reduces to the classical Bayesian MMSE estimator\cite{ain2025,Ozdogan2019,wang2024}:
\[
\bm{\hat{h}}_{k,l}=\bm{\bar{h}}_{k,l}e^{j\theta_{k,l}}+\sqrt{\eta_k}\bm{R}_{k,l}{\bm{\Psi}_{t_k,l}^{'}}\left(\bm{y}_{t_k,l}^{\mathrm{pilot}}-\bm{\bar{y}}_{t_k,l}^{'}\right),
\]
with $\bm{\Psi}_{t_k,l}^{'}=(\sum_{i\in\mathcal{P}_k}\eta_i\tau_p\bm{R}_{i,l}+\sigma^2\bm{I}_N)^{-1}$ and $\bm{\bar{y}}_{t_k,l}^{'}=\sum_{i\in\mathcal{P}_k}\sqrt{\eta_i}\tau_p\bm{\bar{h}}_{i,l}e^{j\theta_{i,l}}$.

In such case, $\bm{\hat{h}}_{k,l}$ and $\bm{\xi}_{k,l}$ are independent random vectors satisfying
$ \mathbb{E}[\bm{\hat{h}}_{k,l}\mid{\theta}_{l}]=\bm{\bar{h}}_{k,l}e^{j\theta_{k,l}},$
$ \mathbb{V}[\bm{\hat{h}}_{k,l}\mid{\theta}_{l}]=\eta_k\tau_p\bm{R}_{k,l}{\bm{\Psi}^{'}_{t_k,l}}\bm{R}_{k,l},$
$ \mathbb{E}[\bm{\xi}_{k,l}\mid{\theta}_{l}]=\bm{0}$ and $ \mathbb{V}[\bm{\xi}_{k,l}\mid{\theta}_{l}]=\bm{R}_{k,l}-\eta_k\tau_p\bm{R}_{k,l}{\bm{\Psi}^{'}_{t_k,l}}\bm{R}_{k,l}$.
\end{corollary}

\begin{corollary}{(Completely unknown phase)}\label{cor:unknown}
For $\varepsilon_{k,l}\sim\mathcal{U}[-\delta,\delta]$, as $\delta\to \pi$ (i.e., $\rho_{\mathrm{error}}\to 0$), the linear MMSE estimator in Proposition~\ref{prop:LMMSE} coincides with the zero-mean model estimate considered in~\cite{Ozdogan2019,wang2021}:
\[
\bm{\hat{h}}_{k,l}=\sqrt{\eta_k}\bm{\Sigma}_{k,l}^{'}{\bm{\Psi}_{t_k,l}^{''}}\bm{y}_{t_k,l}^{\mathrm{pilot}},
\]
where $\bm{\Sigma}_{k,l}^{'}=\bm{\bar{h}}_{k,l}\bm{\bar{h}}_{k,l}^\mathsf{H}+\bm{R}_{k,l}$ and  $\bm{\Psi}_{t_k,l}^{''}=(\sum_{i\in\mathcal{P}_k}\eta_i\tau_p
\bm{\Sigma}_{k,l}^{'} +\sigma^2\bm{I}_N)^{-1}$.

In such case, $\bm{\hat{h}}_{k,l}$ and $\bm{\xi}_{k,l}$ are uncorrelated random vectors satisfying $\mathbb{E}[\bm{\hat{h}}_{k,l}]=\bm{0}$, $\mathbb{V}[\bm{\hat{h}}_{k,l}]=\eta_k\tau_p\bm{\Sigma}_{k,l}^{'}\bm{\Psi}_{t_k,l}^{''}\bm{\Sigma}_{k,l}^{'}$, $\mathbb{E}[\bm{\xi}_{k,l}]=\bm{0}$ and $\mathbb{V}[\bm{\xi}_{k,l}]=\bm{\Sigma}_{k,l}^{'}-\eta_k\tau_p\bm{\Sigma}_{k,l}^{'}\bm{\Psi}_{t_k,l}^{''}\bm{\Sigma}_{k,l}^{'}$.
\end{corollary}

\section{Data Detection}
This section presents the data detection framework under imperfect phase tracking. We begin with the uplink signal model, followed by the beamforming formulations under both centralized and distributed architectures derived by introducing a virtual uplink model, and conclude with the ergodic rate analysis.

\subsection{Uplink Signal Model}

The uplink signal received at AP \( l \) is given by 
\begin{equation}\label{eq:ul_signal}
    \bm{y}_l = \sum_{k=1}^{K} \sqrt{p_k}\bm{h}_{k,l} s_k + \bm{n}_l,
\end{equation}
where $s_k\sim\mathcal{CN}(0,1)$ denotes the independent data-bearing signal transmitted by user $k$ with power $p_k \in \mathbb{R}_+$, and $\bm{n}_l \sim \mathcal{CN}(0,\sigma^2\bm{I}_N)$ represents the additive white Gaussian noise at AP $l$. The decoder computes an estimate of $s_k$ by coherently combining the received signals from all APs as
\[ \hat{s}_k = \sum_{l=1}^L \bm{v}_{k,l}^{\mathsf{H}} \bm{D}_{k,l}\bm{y}_l= \bm{v}_k^{\mathsf{H}} \bm{D}_{k}\bm{y}, \] 
where $\bm{v}_{k} = [\bm{v}_{k,1}^{\mathsf{T}},\dots,\bm{v}_{k,L}^{\mathsf{T}}]^{\mathsf{T}}$ is a $\mathbb{C}^{LN}$-valued random variable that denotes the beamformer across all APs used to detect $s_k$ and $\bm{y}=[\bm{y}_1^\mathsf{T},\dots,\bm{y}_L^\mathsf{T}]^\mathsf{T}\in \mathbb{C}^{LN}$ is the collective uplink signal of the network. Let $\mathcal{L}_k$ denotes the set of APs serving user~$k$, then the selection matrix $\bm{D}_{k,l}=\bm{I}_N$ if $l \in \mathcal{L}_k$ and $\bm{D}_{k,l}=\bm{0}_N$ otherwise. Accordingly, $\bm{D}_{k}=\mathrm{diag}(\bm{D}_{k,1},\dots,\bm{D}_{k,L})$ selects the serving APs for user~$k$.

\subsection{General MMSE beamforming}
We consider the general MMSE framework from~\cite{Miretti2021TeamMP, miretti2025joint} to design the beamformers. In this framework, the beamformers are explicitly modeled and denoted as functions of random variables. This is particularly important for formulating beamforming optimization problems under non-trivial CSI structures. Similar to~\cite{Miretti2021TeamMP, miretti2025joint}, we introduce the following notation:

Given a random quantity \(A\) (e.g., a random vector or matrix), we denote by \(\mathcal{F}^{n}(A)\) the set of \(\mathbb{C}^{n}\)-valued measurable functions of \(A\); that is, random vectors obtained as functions of $A$.

For each user $k$, we then focus on MMSE-optimal beamformers of the form 
\begin{equation}\label{eq:MSE}
\bm{v}_k \in \arg \min_{\bm{u}\in\mathcal{V}_k}\quad \mathbb{E}\left[\left|s_k-\bm{u}^{\mathsf{H}} \bm{D}_{k}\bm{y}\right|^2\right],
\end{equation}
where \(\mathcal{V}_k \subseteq \mathcal{F}^{LN}(\bm{\hat{H}},\bm{\hat\theta})\) is a subset specifying which measurable functions of the available CSI \((\bm{\hat{H}},\bm{\hat\theta})\) can be used to construct the beamformer. 

In the centralized case, \(\mathcal{V}_k\) may coincide with the full set \(\mathcal{F}^{LN}(\bm{\hat{H}},\bm{\hat\theta})\), while in the distributed case, \(\mathcal{V}_k\) may restrict the feasible beamformers to functions of local CSI \(\bm{\hat{H}}_l\) and side information, including \(\bm{\hat\theta}\). For a rigorous formalization of these constraint sets, we refer the reader to~\cite{Miretti2021TeamMP,miretti2025joint}. 

Equation~\eqref{eq:MSE} admits closed-form optimal beamformers when the channel $\bm{H}$ and the channel estimate $\bm{\hat{H}}$ are conditionally jointly complex Gaussian given $\bm{\hat\theta}$, i.e., under the conditions of Corollary~\ref{cor:perfect}, as shown in~\cite[Prop.~8]{miretti2025joint}. However, under imperfect phase knowledge, the above condition is not met in general. This leads to non-trivial expectations such as $\mathbb{E}[\bm{\xi}_{k,l}\mid\bm{\hat{h}}_{k,l}]$, which generally lack simple analytical expressions.

\subsection{Virtual Uplink Model}
To circumvent the above difficulty, we introduce a \emph{virtual uplink model} in which the true channel $\bm{h}_{k,l}$ is replaced by its estimate $\bm{\hat{h}}_{k,l}$, while the noise covariance is augmented with the estimation error covariance. Specifically,
\begin{equation}
    \bm{y}_l^{\mathrm{vir}} = \sum_{k=1}^{K} \sqrt{p_k}\bm{\hat h}_{k,l} s_k + \bm{n}_l^{\mathrm{vir}},
\quad \bm{n}_l^{\mathrm{vir}}\sim \mathcal{CN}(0, \sigma^2\bm{I}_N + \bm{Z}_l),
\end{equation}
where $\bm{Z}_l= \sum_{i=1}^K p_i \bm{C}_{i,l}$ and $\bm{C}_{i,l}$ is defined in \eqref{eq:errorcovariance}. In this virtual model, $\bm{n}^{\mathrm{vir}}_l$ is independent of $\hat{\bm{H}}$, $\bm{\hat\theta}$ and all data symbols, thereby creating a structure where MMSE optimization becomes tractable using the tools introduced in~\cite{Miretti2021TeamMP, miretti2025joint}.

Under this model, the virtual optimal beamformers in the Team MMSE sense~\cite{Miretti2021TeamMP} should satisfy the following conditions 
\begin{equation}\label{eq:MSE_conditional}
(\forall k \in\mathcal{K})\quad\bm{v}_k \in\arg \min_{\bm{u}\in\mathcal{V}_k}\quad \mathbb{E}\left[\left|s_k-\bm{u}^{\mathsf{H}} \bm{D}_{k}\bm{y}^{\mathrm{vir}}\right|^2 \right].
\end{equation}

\begin{remark}
We emphasize that the assumed virtual model is not the true model but an approximation introduced for optimization of the beamformers under phase uncertainty. While the resulting beamformers may not coincide with the globally optimal MMSE solution for the true non-Gaussian channel, they constitute a mathematically rigorous and practically implementable design under imperfect phase knowledge.
\end{remark}

\subsection{Centralized Beamforming}\label{sec:centralized}
In the centralized setting, the CPU has access to the global imperfect CSI \(\hat{\bm{H}}\) and the global phase estimates \(\hat{\boldsymbol{\theta}}\). 
Following~\cite{Miretti2021TeamMP,miretti2025joint}, we model this case by letting \(\mathcal{V}_k = \mathcal{F}^{LN}(\bm{\hat{H}}, \bm{\hat{\theta}})\) in \eqref{eq:MSE_conditional}. In this case, \eqref{eq:MSE_conditional} can be decomposed into disjoint conditional MMSE problems for each realization of $(\hat{\bm{H}},\hat{\bm{\theta}})$, whose point-wise solution readily gives the overall solution
\begin{equation}\label{eq:mmse}
\bm{v}_{k}^{\mathrm{cent}} 
=
\big(
\bm{D}_k \hat{\bm{H}} \bm{P} \hat{\bm{H}}^{\mathsf{H}} \bm{D}_k
+ \bm{D}_k \bm{Z} \bm{D}_k
+ \sigma^2 \bm{I}_{LN}
\big)^{-1}
\bm{D}_k \hat{\bm{H}} \bm{P}^{\frac{1}{2}} \bm{e}_k,
\end{equation}
where \(\bm{P} = \mathrm{diag}(p_1, \dots, p_K)\), \(\bm{e}_k\) is the \(k\)-th column of \(\bm{I}_{K}\), \(\bm{Z} = \mathrm{diag}(\bm{Z}_1, \dots, \bm{Z}_L)\), and \(\bm{D}_{k} = \mathrm{diag}(\bm{D}_{k,1}, \dots, \bm{D}_{k,L})\) selects the serving APs for user~\(k\). 
This expression generalizes the standard MMSE beamformer by embedding the impact of phase estimation errors through the covariance matrices \(\bm{Z}_l\). Note that, although derived by assuming knowledge of the global channel estimate $\bm{\hat H}$, \eqref{eq:mmse} can be calculated by using only the cluster-wise channel estimate $\bm{D}_k\bm{\hat H}$.

\subsection{Distributed Beamforming}\label{sec:distributed}
We now consider a distributed architecture where each AP \(l \in \mathcal{L}\) has access only to its local imperfect CSI \(\hat{\bm{H}}_l\), the global long-term statistics, and the global phase estimates \(\hat{\boldsymbol{\theta}}\). 
Following the Team MMSE framework~\cite{Miretti2021TeamMP}, specialized to the LoS case with perfect phase knowledge in~\cite{ain2025}, we compute distributed beamformers that rely on \(\hat{\boldsymbol{\theta}}\) rather than the true LoS phases. 

For this case, we assume each local beamformer satisfies \(\bm{v}_{k,l} \in \mathcal{F}^{N}(\bm{\hat{H}}_l, \bm{\hat{\theta}})\), and hence 
$\mathcal{V}_k \in \prod_{l=1}^{L} \mathcal{F}^{N}(\bm{\hat{H}}_l, \bm{\hat{\theta}})$ in \eqref{eq:MSE_conditional}. 
The distributed beamforming design can therefore be characterized through a set of \(L\) coupled optimality conditions—one for each AP. 
For every user \(k \in \mathcal{K}\), the solution to \eqref{eq:MSE_conditional} under the considered constraint \(\mathcal{V}_k\) is given by the unique beamformer \((\bm{v}_{k,1}, \dots, \bm{v}_{k,L})\) satisfying $(\forall k \in\mathcal{K})(\forall l \in\mathcal{L}_k)$
% \begin{equation}
% \arg \min_{\bm{u} \in \mathcal{V}_k^{\mathrm{dist}}}
% \mathbb{E}\!\left[
% \Big|
% s_k
% - \bm{u}^{\mathsf{H}} \bm{D}_{k,l} \bm{y}_l^{\mathrm{vir}}
% - \sum_{j \in \mathcal{L}\setminus \{l\}} \bm{v}_{k,j}^{\mathsf{H}} \bm{D}_{k,j} \bm{y}_j^{\mathrm{vir}}
% \Big|^2
% \right].
% \label{eq:teammmse_feas}
% \end{equation}

\begin{equation}
\bm{v}_{k,l}^{\mathrm{dist}}=\bm{V}_{l}\left(\bm{e}_k-\sum_{j \in \mathcal{L}_k\setminus \{l\}} \bm{P}^{\frac{1}{2}} \mathbb{E}[ \hat{\bm{H}}_j^{\mathsf{H}} \bm{v}_{k,j} \mid \hat{\bm{H}}_l,\hat{\boldsymbol{\theta}}]\right),
\label{eq:teammmse_feas}
\end{equation}
where
\begin{equation}\label{eq:lmmse_T}
\bm{V}_{l}=(\hat{\bm{H}}_l\bm{P}\hat{\bm{H}}_l^{\mathsf{H}}+\bm{Z}_l+\sigma^2\bm{I}_{N})^{-1}\hat{\bm{H}}_l\bm{P}^{\frac{1}{2}},\end{equation}
is the local MMSE beamformer at AP $l$. The resulting beamforming vector can then be expressed in closed form as
\begin{equation}\label{eq:tmmse}
(\forall k \in\mathcal{K})(\forall l \in \mathcal{L}_k),\quad\bm{v}_{k,l}^{\mathrm{dist}} = \bm{V}_{l}\bm{c}_{k,l},
\end{equation}
where $\bm{c}_{k,l}\in\mathcal{F}^K(\hat{\theta})$ is a second-stage combining vector, obtained by letting $(\forall l \in \mathcal{L})~\bm{\Pi}_l= \bm{P}^{\frac{1}{2}}\mathbb{E}\{ \hat{\bm{H}}_l^{\mathsf{H}} \bm{V}_l \mid \hat{\boldsymbol{\theta}} \}$ and solving the following system of linear equations~\cite[Thm.~4]{Miretti2021TeamMP}($\forall k \in \mathcal{K}$):

\begin{equation}\label{eq:lin_syst}
\begin{cases}
    \bm{c}_{k,l} + \sum\limits_{j \in \mathcal{L}_k \setminus \{l\}} \bm{\Pi}_j \bm{c}_{k,j} = \bm{e}_k, & \forall l \in \mathcal{L}_k,\\
    \bm{c}_{k,l} = \bm{0}_{K \times 1}, & \text{otherwise}.
\end{cases}
\end{equation}

\begin{remark}
The centralized \eqref{eq:mmse} and distributed \eqref{eq:tmmse} beamformers are constructed solely from second-order statistics of the estimated channels, with phase estimation errors embedded in $\bm{Z}_l$. This yields tractable and practically implementable designs under imperfect LoS phase tracking. Their performance is evaluated by applying them to the true received signal in \eqref{eq:ul_signal}, ensuring a rigorous and fair assessment under realistic LoS phase uncertainty.
\end{remark}

\subsection{Ergodic Rates}\label{sec:ergodic}
Under imperfect phase tracking, both the estimated phase $\hat{\theta}_{k,l}$ and the true phase $\theta_{k,l}$ are random quantities that are statistically related through the phase estimation error. To provide a comprehensive performance assessment, we evaluate the system using two widely used bounds on the achievable spectral efficiency (SE)- namely the lower bound based on use-and-then-forget (UatF)~\cite{Demir2021} decoding and the upper bound based on the optimistic ergodic rate (OER), introduced as a genie-aided bound in~\cite{Demir2021}. These bounds remain valid under imperfect phase tracking, since the expectations in these bounds implicitly average over all sources of randomness, including small-scale fading and LoS phase estimates.

\paragraph{Use-and-then-forget (UatF) Bound}
The UatF bound models the case where instantaneous CSI is exploited for beamforming but not for decoding. The ergodic SE $R_k$ of user $k$ is lower bounded by~\cite{Demir2021}
\begin{equation}\label{eq:se_uatf}
 R_k \geq R_k^{\mathsf{uatf}} =  \log \left(1 + \text{SINR}^{\mathrm{uatf}}_k \right),
\end{equation}
where the effective SINR is given by
\begin{equation*}
\text{SINR}^{\mathrm{uatf}}_k =
\frac{\left| \mathbb{E}\big[ g_{kk} \big] \right|^2}
{\mathbb{E}\big[\sum_{i=1}^K  \left| g_{ki} \right|^2 \big]  
-  \big| \mathbb{E}[ g_{kk}] \big|^2 +\mathbb{E}\big[ \sigma^2 \| \bm{D}_k \bm{v}_k \|^2  \big]},
\end{equation*}
and $\bm{g}_{ki} = \sqrt{p_i}\bm{v}^{\mathsf{H}}_k \bm{D}_k \bm{h}_i$ is the effective channel. 

\paragraph{Optimistic ergodic rate (OER) Bound}
To benchmark the performance limits, we consider the OER bound, which assumes a `genie-aided' decoder that has perfect knowledge of the instantaneous effective channel (including the realization of the phase error), while treating interference and noise as uncorrelated Gaussian noise. The ergodic SE $R_k$ of user $k$ is upper bounded by~\cite{Demir2021}
\begin{equation}\label{eq:se_oer}
 R_k \leq R_k^{\mathsf{oer}} = \mathbb{E}\!\left[ \log \left(1 + \text{SINR}^{\mathsf{oer}}_k \right) \right],
\end{equation}
with the SINR given by \begin{equation*}
\text{SINR}^{\mathrm{oer}}_k =
\frac{ \left| g_{kk} \right|^2}
{\sum\limits_{i\in\mathcal{K}\setminus \{k\}} \left| g_{ki} \right|^2 + \sigma^2 \| \bm{D}_k \bm{v}_k \|^2 }.
\end{equation*}

\begin{remark}\label{rem:ergodic}
Since the expectations in \eqref{eq:se_uatf} and \eqref{eq:se_oer} are taken over all random effects, including residual LoS phase estimation errors, the ergodic bounds directly reflect the performance degradation due to imperfect phase tracking without requiring any modification to their standard analytical form.
\end{remark}

\section{Numerical Results}\label{sec:results}

\begin{figure*}[t!]
\centering
\subfloat[]
{\includegraphics[width=0.33\textwidth]{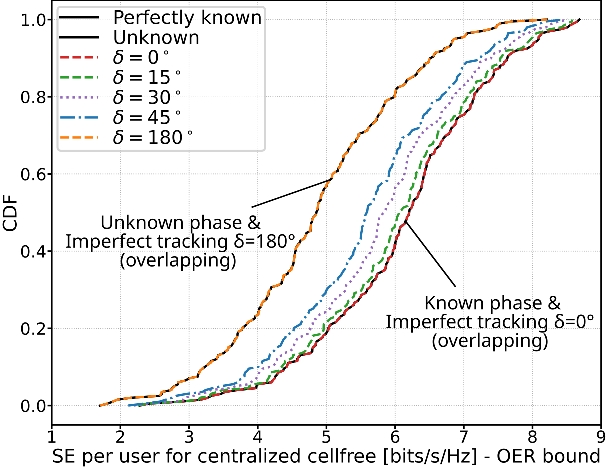}\label{fig:cdf_mmse_oer}}
\hfil
\subfloat[]
{\includegraphics[width=0.33\textwidth]{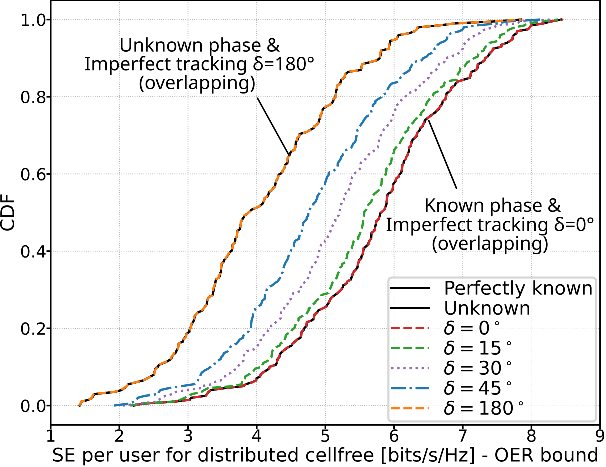}\label{fig:cdf_tmmse_oer}}
\hfil
\subfloat[]
{\includegraphics[width=0.33\textwidth]{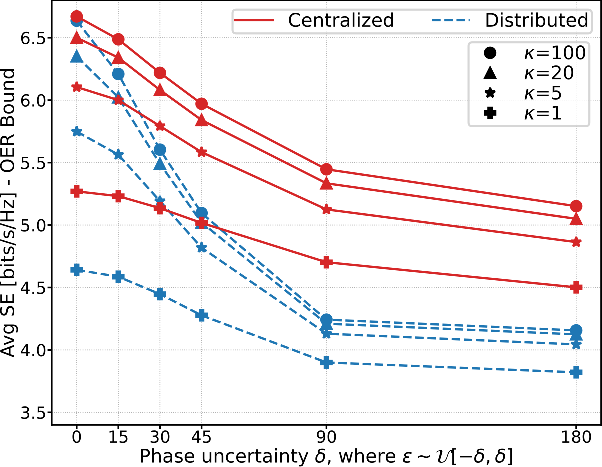}\label{fig:avg_ses_oer}}
\caption{
Comparison of uplink SE obtained using the OER bound~\eqref{eq:se_oer} under different LoS phase-tracking scenarios. 
Figure~\ref{fig:cdf_mmse_oer} shows the CDF of the per-user SE for centralized cell-free beamforming, while Figure~\ref{fig:cdf_tmmse_oer} shows the corresponding CDF for distributed cell-free beamforming. 
In both (a) and (b), the black solid curves represent two benchmark cases from the literature: the optimistic case with perfectly known LoS phases (Cor.~\ref{cor:perfect}) and the pessimistic case with completely unknown phases (Cor.~\ref{cor:unknown}). 
The dotted curves correspond to the proposed model in Proposition~\ref{prop:LMMSE}, where the LoS phase is partially tracked with an error $\varepsilon \sim \mathcal{U}[-\delta,\delta]$. 
Results are shown for $\delta \in \{0^{\circ},15^{\circ},30^{\circ},45^{\circ},180^{\circ}\}$ with a fixed Rician factor $\kappa=5$ for all channels. 
Figure~\ref{fig:avg_ses_oer} shows the average network SE versus phase uncertainty for centralized and distributed beamforming at different Rician factors $\kappa \in \{1,5,20,100\}$ and $\delta \in \{0^{\circ},15^{\circ},30^{\circ},45^{\circ},90^{\circ},180^{\circ}\}$.
}
\label{fig:results_v1}
\end{figure*}
\begin{figure*}[t!]
\centering
\subfloat[]
{\includegraphics[width=0.33\textwidth]{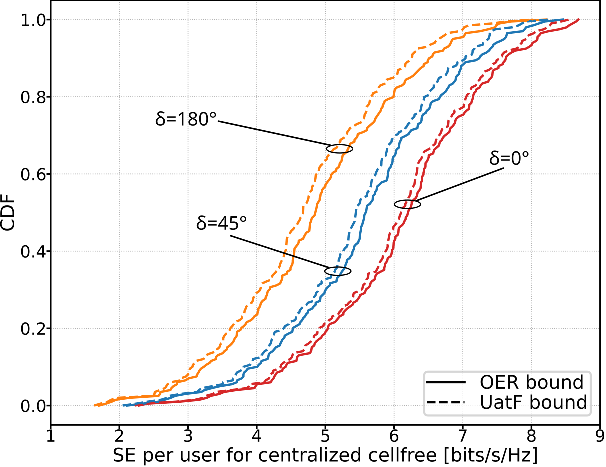}\label{fig:cdf_mmse_both}}
\hfil
\subfloat[]
{\includegraphics[width=0.33\textwidth]{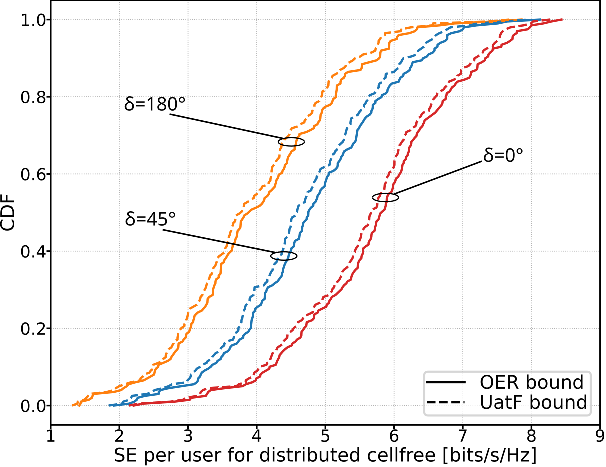}\label{fig:cdf_tmmse_both}}
\hfil
\subfloat[]
{\includegraphics[width=0.33\textwidth]{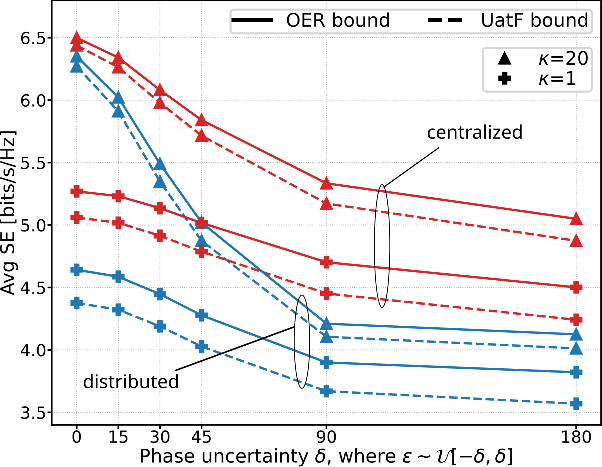}\label{fig:avg_ses_both}}
\caption{
Comparison of upper (OER) and lower (UatF) bounds on the achievable SE under imperfect LoS phase tracking for centralized beamforming in Figure~\ref{fig:cdf_mmse_both} and distributed beamforming in Figure~\ref{fig:cdf_tmmse_both}. 
In both (a) and (b), solid curves correspond to the OER bound, while dotted curves represent the UatF bound. 
Results are shown for $\delta \in \{0^{\circ},45^{\circ},180^{\circ}\}$ with a fixed Rician factor $\kappa=5$ for all channels. 
Figure~\ref{fig:avg_ses_both} shows the average network SE versus phase uncertainty for centralized and distributed beamforming at Rician factors $\kappa \in \{1,20\}$, using the OER and UatF bounds (solid and dotted curves, respectively).
}
\label{fig:results_v2}
\end{figure*}

\subsection{Parameters and Setup}
\begin{table}[ht]
\centering
\renewcommand{\arraystretch}{1.2} % Increases row spacing
\setlength{\tabcolsep}{10pt} % Adjusts the space between columns
\begin{tabular}{||c c||} 
\hline
\textbf{Parameter} & \textbf{Value} \\ [1ex]
\hline\hline
Network area & $1000\,\text{m}\,\times 1000\,\text{m}$ \\
\hline
Number of \acp{ap} & $L=100$\\
\hline
Number of UEs & $K=40$\\
\hline
Number of antennas per AP & $N=4$ \\
\hline
Bandwidth & $B=100$ MHz\\
\hline
Carrier frequency & $f_c=5$ GHz\\
\hline
Maximum uplink transmit power & $p_{\mathrm{max}}=100$ mW\\
\hline
Coherence block symbols & $\tau_c=200$\\
\hline
Pilot symbols & $\tau_p=5$\\
\hline
AP-user height difference & $\Delta h=11$ m\\
\hline
Shadow fading deviation & $\sigma_{\mathrm{sf}} = 8$ dB \\
\hline
Antenna spacing & $d=\lambda/2$\\
\hline
\end{tabular}\\
\vspace{1mm} % Add space between the table and the caption
\caption{Simulation parameters}
\label{tab:parameters}
\end{table}

 We consider a cell-free massive MIMO network with $L=100$ \acp{ap}, each equipped with a uniform linear array (ULA) with $N=4$ antennas, jointly serving $K=40$ single-antenna users distributed uniformly at random over a $1000\,\text{m} \times 1000\,\text{m}$ area. A wrap-around technique is applied to emulate an infinite service region. The large-scale fading coefficient between \ac{ap}~$l$ and user~$k$ is modeled as~\cite{3gpp}
\begin{equation}
\beta_{k,l}= 35.4 - 20\log_{10}(f_c)-26\log_{10}\!\left( \tfrac{d_{k,l}}{1\text{m}}\right)+F_{k,l},\quad [\text{dB}],
\end{equation}
where $d_{k,l}$ is the 3D distance, $f_c$ is the carrier frequency, and $F_{k,l}\sim\mathcal{N}(0,\sigma_{\mathrm{sf}}^2)$ represents shadow fading. The small-scale fading follows a spatially correlated Rician model.
Other aspects of the channel modeling, spatial correlation, clustering, and power control follow details in~\cite{ain2025}. For a fixed network experiment, $1000$ Monte Carlo trials are performed, where in each trial the LoS phases are sampled as independent and identically distributed uniform random variables over $[-\pi,\pi]$. The main simulation parameters are summarized in Table~\ref{tab:parameters}, and further details can be found in~\cite{ain2025}.

\subsection{Results and Conclusions}
Figures~\ref{fig:cdf_mmse_oer} and \ref{fig:cdf_tmmse_oer} show the cumulative distribution functions (CDFs) of the uplink SE per-user, computed using the OER bound~\eqref{eq:se_oer}, for centralized (Section~\ref{sec:centralized}) and distributed (Section~\ref{sec:distributed}) beamforming in a cell-free network, respectively. We consider LoS propagation and fix the Rician factor at $\kappa=5$ for all channels. The black curves represent the two benchmarks from the literature: the optimistic case of perfectly known LoS phases (Corollary~\ref{cor:perfect}) and the pessimistic case of completely unknown phases (Corollary~\ref{cor:unknown}). The intermediate cases, corresponding to the imperfect phase tracking captured by our proposed model in Proposition~\ref{prop:LMMSE}, are illustrated by dotted curves for $\delta\in\{0^{\circ},15^{\circ},30^{\circ},45^{\circ},180^{\circ}\}$, where the phase error is $\varepsilon\sim\mathcal{U}[-\delta,\delta]$. The results demonstrate that even partial knowledge of the slowly varying LoS phases provides substantial performance gains compared to the no-phase-knowledge scenario. As $\delta$ increases, the gap to the perfect knowledge benchmark widens; however, moderate errors (e.g., $\delta=15^{\circ}$) already result in distributions close to the ideal case for both architectures. Noticeable degradation appears only for large phase uncertainties, particularly in a distributed setting.

Figure~\ref{fig:avg_ses_oer} illustrates the average network SE as a function of the phase uncertainty, evaluated using the OER bound~\eqref{eq:se_oer}. The results compare centralized and distributed beamforming under different Rician factors $\kappa\in\{1,5,20,100\}$. At high Rician factors (e.g., $\kappa=100$), both designs tend to converge, and the performance gap between centralized and distributed cell-free networks becomes negligible when the LoS phases are accurately known. This observation is consistent with our previous findings~\cite{ain2025}, where we showed that under strong LoS conditions and perfect phase knowledge, distributed beamforming that exploits the statistical channel information can closely approach the centralized benchmark. However, in the presence of phase uncertainty, the achievable rates of both architectures degrade, with the impact becoming more pronounced as $\kappa$ increases. Nevertheless, even coarse or partially accurate phase tracking provides significant performance gains over the no-phase knowledge model, particularly for distributed designs.

In Figure~\ref{fig:results_v2}, we evaluate both $R_{k}^{\text{uatf}}$~\eqref{eq:se_uatf} and $R_{k}^{\text{oer}}$~\eqref{eq:se_oer} under imperfect LoS phase tracking to assess the tightness of the SE bounds. The gap between the two bounds is observed to be small, particularly for $\kappa=20$ in Figure~\ref{fig:avg_ses_both} with $\delta\in\{0^\circ,15^\circ\}$. This behavior indicates that due to the dominant LoS component with an almost perfectly known LoS phase, the effective channel $\bm{g}_{kk}$ behaves nearly deterministically, meaning that the lack of instantaneous CSI at the decoder results in a negligible performance loss.
% Conversely, a larger gap for $\delta\in[90^\circ,180^\circ]$ suggests that the effective channel fluctuates significantly due to residual phase errors and other sources of randomness.

Overall, the results demonstrate that partial knowledge of LoS phases remains highly beneficial for both centralized and distributed beamforming, with centralized beamforming exhibiting greater robustness under severe phase errors. These findings underscore the importance of phase-aware channel estimation and beamforming strategies for practical 6G cell-free massive MIMO deployments.

% ==== Switch to one column for derivation 
\section{Appendix}
\small
This appendix provides detailed proofs of Lemma 1 and Proposition 1, which were omitted from the main text.
\subsection{Proof of Lemma 1}\label{app:lemma}
We derive the first and second moments of the channel for a specific AP-user pair. For brevity, we drop the user and AP indices $\{k,l\}$ and denote the phase error by $\epsilon \sim \mathcal{U}[-\delta, \delta]$. 
The characteristic function of the uniform phase error is calculated as
\begin{align}
    \rho_{\text{error}} \triangleq \mathbb{E}[e^{j\epsilon}] &= \frac{1}{2\delta} \int_{-\delta}^{\delta} e^{jx}dx \nonumber 
    = \begin{cases} 1, & \delta = 0, \\ \frac{\sin(\delta)}{\delta}, & 0 < \delta \leq \pi. \end{cases}
    \label{eq:rho_def}
\end{align}
Note that for the symmetric interval $[-\delta, \delta]$, $\rho_{\text{error}}$ is real-valued.

\textit{1) Conditional Expectation:} 
Substituting $\theta = \hat{\theta} + \epsilon$ into the channel model \eqref{eq:channel}, we have $\mathbf{h} = \bar{\mathbf{h}}e^{j(\hat{\theta}+\epsilon)} + \tilde{\mathbf{h}}$. 
The terms $\bar{\mathbf{h}}$ and $e^{j\hat{\theta}}$ (conditioned on the estimate $\hat{\theta}$) are deterministic, while $\tilde{\mathbf{h}}$ remains zero-mean and independent of the phase error. 
Taking the conditional expectation yields
\begin{align}
    \boldsymbol{\mu} \triangleq\mathbb{E}[\mathbf{h} \mid \hat{\theta}] 
    &= \bar{\mathbf{h}} e^{j\hat{\theta}} \mathbb{E}[e^{j\epsilon}] + \mathbb{E}[\tilde{\mathbf{h}}] \nonumber = \bar{\mathbf{h}} e^{j\hat{\theta}} \rho_{\text{error}}.
\end{align}

\textit{2) Conditional Covariance:} 
The conditional covariance matrix is defined as
\begin{equation}\label{app:channel_variance}
\boldsymbol{\Sigma} = \mathbb{E}[(\mathbf{h}-\boldsymbol{\mu})(\mathbf{h}-\boldsymbol{\mu})^\mathsf{H} \mid \hat{\theta}], \end{equation}
where the deviation from the expectation is given by $\mathbf{h} - \boldsymbol{\mu} = \bar{\mathbf{h}} e^{j\hat{\theta}} (e^{j\epsilon} - \rho_{\text{error}}) + \tilde{\mathbf{h}}.$ 
Since $\epsilon$ and $\tilde{\mathbf{h}}$ are independent and $\tilde{\mathbf{h}}$ is zero-mean, the cross-terms in the expectation vanish. Equation \eqref{app:channel_variance} simplifies to
\begin{align}
    \boldsymbol{\Sigma} &= \bar{\mathbf{h}}\bar{\mathbf{h}}^\mathsf{H} \mathbb{E}[|e^{j\epsilon} - \rho_{\text{error}}|^2] + \mathbb{E}[\tilde{\mathbf{h}}\tilde{\mathbf{h}}^\mathsf{H}] \nonumber \\
    &= \bar{\mathbf{h}}\bar{\mathbf{h}}^\mathsf{H} (1 - \rho_{\text{error}}^2) + \mathbf{R} \quad \text{(using } |e^{j\epsilon}|^2 = 1\text{)},
\end{align}
which concludes the proof. 
\hfill $\blacksquare$

\subsection{Proof of Proposition 1}\label{app:prop}
We derive the conditionally linear MMSE estimator for a specific AP-user pair. For clarity, we omit the user index $k$, AP index $l$, and pilot index $t_k$. Let $\mathbf{h}$ denote the channel of the user of interest, $\hat{\theta}$ the phase estimate, and $\mathbf{y}$ the observed signal for the assigned pilot.

\textit{1) Preliminaries:}
From \eqref{eq:ul_pilot_signal}, the received pilot signal from the set of pilot-sharing users $\mathcal{P}$ is given by $
\mathbf{y} = \sum_{i \in \mathcal{P}} \sqrt{\eta_i} \tau \mathbf{h}_{i} + \mathbf{n},
$
where $\mathbf{n} \sim \mathcal{CN}(\mathbf{0}, \sigma^2 \tau \mathbf{I}_N)$. 
Using the result $\mathbb{E}[\mathbf{h}_i | \hat{\theta}_i]=\boldsymbol{\mu}_i$ from the proof of Lemma 1, the conditional expectation of the observed signal is
\begin{equation}
    \bar{\mathbf{y}} \triangleq \mathbb{E}[\mathbf{y} | \hat{\boldsymbol{\theta}}] = \sum_{i \in \mathcal{P}} \sqrt{\eta_i} \tau \boldsymbol{\mu}_i.
\end{equation}

The conditional cross-covariance between the target channel $\mathbf{h}$ and the observation $\mathbf{y}$ is $\mathbf{C}_{hy} = \mathbb{E}[(\mathbf{h}-\boldsymbol{\mu})(\mathbf{y}-\bar{\mathbf{y}})^\mathsf{H} \mid \hat{\boldsymbol{\theta}}].$
Exploiting the independence of user channels and receiver noise, it simplifies to
\begin{equation} \mathbf{C}_{hy} = \mathbb{E}[(\mathbf{h}-\boldsymbol{\mu})(\sqrt{\eta}\tau (\mathbf{h}-\boldsymbol{\mu}))^\mathsf{H} \mid \hat{\boldsymbol{\theta}}] = \sqrt{\eta}\tau \boldsymbol{\Sigma},
\end{equation}
where $\eta$ and $\boldsymbol{\Sigma}$ refer to the parameters of the user of interest.

Similarly, exploiting the independence of user channels and receiver noise, the conditional covariance matrix of the observed signal is derived as
\begin{align}
    \mathbf{C}_{yy}= \mathbb{E}[(\mathbf{y}-\bar{\mathbf{y}})(\mathbf{y}-\bar{\mathbf{y}})^\mathsf{H} \mid \hat{\boldsymbol{\theta}}]= \sum_{i \in \mathcal{P}} \eta_i \tau^2 \boldsymbol{\Sigma}_{i} + \sigma^2 \tau \mathbf{I}_N,
\end{align}
where $\boldsymbol{\Sigma}_i$ is the conditional channel covariance of user $i$ defined in Lemma 1.

\textit{2) Estimator Derivation:}
Substituting the moments derived above into the Definition~\eqref{def:lmmse} of conditionally linear MMSE estimator gives
\begin{equation}
    \hat{\mathbf{h}} = \boldsymbol{\mu} + \mathbf{C}_{hy} \mathbf{C}_{yy}^{-1} (\mathbf{y} - \bar{\mathbf{y}}).
\end{equation}
To recover the expression in Proposition 1, $\tau$ can be factored out as follows
\begin{align}
    \mathbf{C}_{hy} \mathbf{C}_{yy}^{-1} &= (\sqrt{\eta} \tau \boldsymbol{\Sigma}) \big( \tau \big[ \sum_{i \in \mathcal{P}} \eta_i \tau \boldsymbol{\Sigma}_{i} + \sigma^2 \mathbf{I}_N \big] \big)^{-1} \nonumber \\
    &= \sqrt{\eta} \boldsymbol{\Sigma} \underbrace{\big( \sum_{i \in \mathcal{P}} \eta_i \tau \boldsymbol{\Sigma}_{i} + \sigma^2 \mathbf{I}_N \big)^{-1}}_{\triangleq \boldsymbol{\Psi}}.
\end{align}
\begin{align}\text{Thus,}\quad\hat{\mathbf{h}} =\boldsymbol{\mu} + \sqrt{\eta} \boldsymbol{\Sigma} \boldsymbol{\Psi} (\mathbf{y} - \bar{\mathbf{y}}).\end{align}

Finally, the error covariance matrix for a linear MMSE estimator is given by $\mathbf{C} = \boldsymbol{\Sigma} - \mathbf{C}_{hy} \mathbf{C}_{yy}^{-1} \mathbf{C}_{hy}^\mathsf{H}$ \cite{kay_1993}, which simplifies to \begin{align}\mathbf{C} = \boldsymbol{\Sigma} - \eta \tau \boldsymbol{\Sigma} \boldsymbol{\Psi} \boldsymbol{\Sigma}.\end{align}\\
This concludes the proof. 
\hfill $\blacksquare$

\ifCLASSOPTIONcaptionsoff
  \newpage
\fi

\bibliographystyle{IEEEbib}
\bibliography{bibtex/bib/references}

\end{document}

%% file: macros.tex
\usepackage{cite}
\ifCLASSINFOpdf
  \usepackage[pdftex]{graphicx}
\else
   \usepackage[dvips]{graphicx}
\fi
\usepackage{amsmath,bm}
\usepackage{amssymb}

\usepackage{mathtools}
\ifCLASSOPTIONcompsoc
  \usepackage[caption=false,font=normalsize,labelfont=sf,textfont=sf]{subfig}
\else
  \usepackage[caption=false,font=footnotesize]{subfig}
\fi

\usepackage{stfloats}
\usepackage{url}
\usepackage{multirow}
\usepackage{graphicx}

\usepackage{svg}
\usepackage[printonlyused,nohyperlinks,nolist]{acronym}
\usepackage{algorithm}
\usepackage{algpseudocode}

\usepackage{dsfont}

\usepackage{siunitx}
\usepackage{import}
\usepackage{color}

\DeclareUnicodeCharacter{2212}{-}

\def\BibTeX{{\rm B\kern-.05em{\sc i\kern-.025em b}\kern-.08em
    T\kern-.1667em\lower.7ex\hbox{E}\kern-.125emX}}

\usepackage{amsthm}
\newtheorem{remark}{Remark}
\newtheorem{proposition}{Proposition}
\newtheorem{lemma}{Lemma}
\newtheorem{definition}{Definition}
\newtheorem{corollary}{Corollary}

\usepackage{scalefnt}

\usepackage[top=0.75in, bottom=1in, left=0.625in, right=0.625in]{geometry}
\usepackage{mathtools} 

%% file: acronyms.tex
%------------------------------------------------------------------
% Acronyms
% use:  \ac{BS}         to use a acronym (first time = full name, then only acronym)
%       \acp{BS}        use the plural
%       \acf{BS}        print the full name and ignore previous declaration
%       \acs{BS}        Use the acronym, even before the first corresponding \ac command
%       \acl{acronym}   Expand the acronym without using the acronym itself.
%       \acresetall     Reset all acronyms (useful after abstract)
%------------------------------------------------------------------
\begin{acronym}
    \acro{los}[LoS]{line-of-sight}
    \acro{ul}[UL]{uplink}
    \acro{dl}[DL]{downlink}
    \acro{nlos}[NLoS]{non-line-of-sight}
    \acro{ap}[AP]{access point}
    \acro{ue}[UE]{user equipment}
    \acro{mimo}[MIMO]{multiple-input multiple-output}
    \acro{mmimo}[mMIMO]{massive multiple-input multiple-output}
    \acro{csi}[CSI]{channel state information}
    \acro{sinr}[SINR]{signal to interference and noise ratio}
    \acro{snr}[SNR]{signal-to-noise ratio}
    \acro{mmse}[MMSE]{minimum mean square error}
    \acro{lmmse}[LMMSE]{local \acl{mmse}}
    \acro{tmmse}[TMMSE]{team \acl{mmse}}
    \acro{ltmmse}[LTMMSE]{local team \acl{mmse}}
    % \acro{mmse}[MMSE]{minimum mean square error}
    % \acro{lmmse}[LMMSE]{local MMSE}
    % \acro{tmmse}[TMMSE]{team minimum mean square error}
    % \acro{ltmmse}[LTMMSE]{local team MMSE}

    \acro{emmse}[E-MMSE]{Element-wise MMSE}
    \acro{mr}[MR]{maximum-ratio}
    \acro{zf}[ZF]{zero-forcing}
    \acro{ls}[LS]{least square}
    \acro{mse}[MSE]{mean squared error}
    \acro{se}[SE]{spectral efficiency}
    \acro{dcc}[DCC]{dynamic cooperation clustering}
    \acro{lsfd}[LSFD]{large scale fading decoding}
    \acro{sota}[SOTA]{state-of-the-art}
    \acro{comp}[CoMP]{coordinated multipoint}
    \acro{cpu}[CPU]{central processing unit}
    \acro{uatf}[UatF]{use-and-then-forget}
    \acro{cd}[CD]{coherent decoding}
\end{acronym}